\documentclass[useAMS,fleqn, usenatbib, final]{mnras}
\usepackage{lipsum}
\usepackage{multirow}
\usepackage[inline]{enumitem}
\usepackage{import}
\usepackage{tabularx}
\usepackage{glossaries}
\usepackage[titletoc,title]{appendix}
\usepackage{graphicx}
\usepackage{amsmath}
\usepackage{amssymb}
\usepackage[normalem]{ulem}
\usepackage[normalem]{ulem}
\usepackage{amsfonts}
\usepackage{mathtools}
\usepackage{bm}
\usepackage[utf8]{inputenc}
\usepackage[table]{xcolor}
\usepackage{url}
\usepackage[normalem]{ulem}

\newcommand{\paperdoi}{10.5281/zenodo.23504465}
\newcommand{\lisanoisedoi}{10.5281/zenodo.23501674}

\newcommand{\portsmouth}{Institute of Cosmology and Gravitation, University of Portsmouth, Portsmouth, PO1 3FX, UK}
\newcommand{\glasgow}{School of Physics and Astronomy, University of Glasgow, Glasgow, G12 8QQ, UK}

\title[Non-parametric Noise Inference for LISA]{Noisy pipelines \uppercase\expandafter{\romannumeral 2}: Non-parametric Noise Inference for LISA}

\author[Hoy]{
\parbox{\textwidth}{
C.~Hoy$^{1}$\thanks{charlie.hoy@port.ac.uk},
J.~Bayley$^{2}$,
O.~Burke$^{2}$,
L.K.~Nuttall$^{1}$,
}
\vspace{0.2cm}\\
$^1$\portsmouth\\
$^2$\glasgow\\
\vspace{-1.5em}
}
\date{Accepted XX. Received YY; in original form ZZ}
\pubyear{2026}

\begin{document}
\label{firstpage}
\pagerange{\pageref{firstpage}--\pageref{lastpage}}

\maketitle

\begin{abstract}
The Laser Interferometer Space Antenna (LISA) is expected to observe gravitational-wave signals from massive black hole binaries, extreme mass ratio inspirals, and more. Despite the significant scientific potential, accurately extracting these signals from the data will only be possible if LISA's instrumental noise is understood: poor noise modelling will result in lost signals, biased parameter estimates, and incorrect astrophysical insights. Modelling LISA's instrumental noise is more challenging than its ground-based counterpart since the data is highly correlated, meaning that the noise covariance matrix can no longer be assumed diagonal in frequency and in the different data streams. In this work, we present an initial approach to perform non-parametric noise inference for LISA that requires no prior knowledge of the LISA observatory. By fitting the full noise covariance matrix with a combination of splines and inverse Lorentzians, we show that our model can estimate the power and cross spectral densities for LISA from 7 days of simulated noise-only data. We find that the median estimate from our non-parametric model whitens the data according to the expected unit normal distribution. We envision that our non-parametric model can be used as part of the global fit, as well as in combination with modelled approaches to identify, quantify, and constrain the frequency dependence of unknown noise artifacts.
\vspace{1.5em}
\end{abstract}

\section{Introduction}

Similar to existing ground-based gravitational-wave (GW) detectors, the raw data from future space-based detectors, such as the Laser Interferometer Space Antenna~\citep[LISA,][]{LISA2017,LISA:2024hlh}, are expected to be dominated by fundamental, technical\footnote{Fundamental noise sources arise from the physics of the measurement itself, such as thermal and photon shot noise. Technical noise sources arise from imperfections in the detector’s implementation, such as noise from electronics.} and unknown noise sources. Once the dominant technical noise sources are suppressed, making LISA source (or, in part of its band, confusion) limited, data analysis methods, such as searching for signals~\citep[see e.g.][]{Bandopadhyay:2023gkb,Weaving:2023fji,Bandopadhyay:2024nys,CabournDavies:2024hea,Strub:2025dfs,CabournDavies:2026ldd} and inferring their properties~\citep[see e.g.][]{Vecchio:2003tn,Berti:2004bd,Lang:2006bsg,Arun:2008zn,Cornish:2021smq,Katz:2021uax,Littenberg:2023xpl,Klein:2022rbf,Pratten:2022kug,Hoy:2023ndx,Hoy:2024ovd}, rely on being able to statistically differentiate the data from the expected noise distribution~\citep{Finn:1992wt,Veitch:2008ur,Veitch:2014wba,Thrane:2018qnx}. Having a well understood noise model is crucial to ensure accurate astrophysical parameter estimation, concrete discoveries of new physics, and to maximise the scientific return of these space-based missions.

Time delayed interferometry~\citep[TDI,][]{Tinto:1999yr} is a post-processing technique to suppress one of LISA's most significant noise sources: laser frequency noise. This technical noise source is orders of magnitude larger than any expected GW signal, making GW detection challenging. TDI is a technique that combines the data from LISA's three individual spacecraft to form virtual interferometric observables. Numerous observables have been proposed in the past with different underlying assumptions, but the second-generation TDI observables: $X$, $Y$ and $Z$, allow for linearly changing arm lengths, making them the most applicable for realistic LISA scenarios~\citep{armstrong1999time,Estabrook:2000ef,Vallisneri:2005ji,Tinto:2020fcc,tinto2023second}.

Even after TDI has been performed, estimating LISA's noise distribution is more challenging than its ground-based counterpart. This is because unlike the existing LIGO~\citep{LIGOScientific:2014pky}, Virgo~\citep{VIRGO:2014yos}, and KAGRA~\citep{KAGRA:2020tym} GW detectors, the $X$, $Y$ and $Z$ TDI observables are correlated in their noise properties. As such, the cross spectral densities (CSDs) can no longer be assumed zero, and must be modelled alongside the power spectral densities (PSDs). A standard fix is to linearly combine the $X$, $Y$ and $Z$ observables, forming a set of approximately uncorrelated TDI observables for analyses. However, these $A$, $E$ and $T$ channels are only uncorrelated under idealised assumptions: equal arm lengths and identical noise levels in every test mass and optical link~\citep{Prince:2002hp}. It is therefore important to develop methods beyond those used in ground-based GW detectors to estimate LISA's correlated noise distribution.

Several techniques have already been proposed to estimate LISA's noise distribution. These can broadly be divided into modelled, non-parametric and Welch-based approaches. \cite{Katz:2024oqg,deng2025modular} modelled the instrumental noise as the sum of the two fundamental LISA noise sources -- the Test Mass Acceleration (TMA) and Optical Metrology System (OMS) noise -- each with a fixed powerlaw and a free amplitude, together with a five-parameter phenomenological model for the confusion foreground from unresolved Galactic binaries. In contrast, \cite{Littenberg:2023xpl} adapted the non-parametric \textsc{BayesLine} algorithm~\citep{Littenberg:2014oda} to fit the LISA noise with an Akima spline, using a reference noise spectrum that encodes important features in the LISA response. \cite{Baghi:2023qnq,Burke:2026aaa} both used flexible splines that were transformed to the TDI observables through given transfer functions, and \cite{Aimen:2025zxn} proposed using penalised B-splines (P-splines) to model the noise from a parametric guess. \cite{Vajpeyi:2026pzh} later extended this work to model the noise with P-splines assuming no knowledge of the LISA detector. Finally, \cite{Strub:2024kbe} uses a combination of Welch PSDs~\citep{Welch:1967oth} and a spline to fit the LISA noise. 

Since flexible non-parametric models are regularly used throughout GW astronomy, we present a non-parametric noise model for LISA that requires no prior information about the LISA observatory. We show that our model can fit the full noise covariance matrix -- both the PSDs and CSDs for correlated data -- for the $X$, $Y$ and $Z$ TDI observables. Owing to its flexibility, we envision that this model can be used as part of the global fit, as well as in combination with other modelled approaches to identify possible deviations from the fundamental noise sources arising from unknown effects. Further investigations are needed to identify how the model performs on e.g. different data durations~\citep[see e.g.][]{Vajpeyi:2026pzh}, including gaps in the data~\citep[see e.g.][]{Burke:2026rjq}.

This paper is structured as follows: in Section~\ref{sec:noise_sources} we discuss the expected noise sources for the LISA observatory and in Section~\ref{sec:models} we discuss different techniques for modelling the LISA's noise distribution, including the new non-parametric noise model described in this work. In Section~\ref{sec:likelihood} we review the Bayesian methods used to infer the noise model properties and in Section~\ref{sec:results} we discuss our results on simulated LISA data. Finally we conclude in Section~\ref{sec:discussion}.

\section{LISA noise sources} \label{sec:noise_sources}

LISA will consist of three spacecraft, each containing a free falling test mass. Lasers will be relayed between each of the spacecraft, and the distance between free falling test masses will be monitored over time.

LISA's fundamental noises are well known, and their frequency dependence well documented: TMA noise limits the low-frequency sensitivity and follows a powerlaw with displacement PSD $\propto f^{-4}$; OMS noise limits the high-frequency sensitivity and follows a broken powerlaw with a displacement PSD $\sim 1 + (2\mathrm{mHz}/f)^4$~\citep{Robson:2018ifk}. TMA noise arises from stray forces acting on the free-falling test masses inside each spacecraft, such as random, position independent forces and couplings to the noisy motion of the spacecraft shield~\citep{Bortoluzzi:2004cz}. The TMA noise becomes dominant below $\sim 1\,\mathrm{mHz}$, as demonstrated by LISA Pathfinder~\citep{Armano:2018kix}. The high frequency sensitivity is controlled by the OMS noise, which measures the distance of the free-falling test masses with respect to the reference test mass~\citep{bortoluzzi2004science}. The OMS noise is fundamentally limited by photon shot noise. However, for LISA it dominates above a few $m\mathrm{Hz}$, where the TMA noise becomes subdominant~\citep[see e.g.][]{Bayle:2022okx}. Other technical noise sources, such as tilt-to-length coupling noise, clock noise, laser modulation and sideband noise will also affect LISA~\citep{Hartig_2022, PhysRevApplied.14.014030,PhysRevD.103.123027}, but for this work, we only consider OMS and TMA secondary noise sources.

The major noise contribution of the LISA observatory is laser frequency noise. This will be several orders of magnitude larger than any change in arm length due to expected GW. The laser frequency noise can be suppressed by performing an additional post-processing step known as TDI. TDI is a technique that combines data from the three spacecraft into virtual interferometric observables, where the laser frequency noise is suppressed by several orders of magnitude. The second-generation TDI observables, $X$, $Y$ and $Z$, allowed for linearly changing arm lengths, making them the most applicable for realistic LISA scenarios. For this reason, we only consider the $X$, $Y$ and $Z$ variables in this work, but this brings about additional challenges as they are correlated in their noise properties.

\section{LISA noise models} \label{sec:models}

Numerous techniques have been proposed to model the noise for the LISA observatory~\citep{Littenberg:2023xpl,Baghi:2023qnq,Strub:2024kbe,Katz:2024oqg,deng2025modular,Aimen:2025zxn,Burke:2026aaa}. A Welch estimate is the most common technique in astronomy~\citep[see e.g.][]{Allen:2005fk}, where the timeseries data are divided into smaller, overlapping segments and the individual periodograms are averaged to reduce the variance and noise~\citep{Welch:1967oth}. However, a Welch estimate is non-ideal for LISA as the noise will be non-stationary with many overlapping signals. Unlike in ground-based GW detectors, LISA will not have a signal-free segment of data to estimate the noise distribution off-source, and LISA's arm lengths will breathe by a few percent a year. This means that the nulls in the transfer functions in particular will drift with time~\citep{Tinto:2003vj,tinto2014time,Bayle:2022okx,Burke:2026aaa}.

Various modelled approaches have been proposed in the past. In~\cite{Burke:2026aaa}, we proposed a technique that took advantage of the known geometry of the LISA observatory, specifically the light travel times between individual spacecraft. This technique modeled the expected fundamental LISA noises with splines and used the given transfer functions to project the single-link measurements into TDI observables. This assumes that there are no other noise sources beyond the OMS and TMA noise; the splines have the freedom to fit additional noise sources, but the inferred TMA and OMS noise measurements will be biased. Once LISA flies, this underlying assumption may not be valid, especially for long duration data where glitches, and potentially instrumental malfunctions are more likely. For example, although ground-based GW detectors have been built and studied in detail for decades, there still exists unknown noise sources that affect the sensitivity of the instrument especially at low frequencies, see e.g.~\cite{Capote:2024rmo}. Similarly, in more realistic LISA scenarios, the secondary noise spectra become more complex, significantly increasing the the dimensionality of the problem, and number of splines to sample over~\citep{2023PhRvD.108h2004Q,Hartwig:2023pft}. For this reason, there is motivation to develop an unmodeled approach that relaxes these assumptions.

Owing to their flexibility, non-parametric approaches that estimate the noise spectrum are routinely used for ground-based GW astronomy. For example, all GW detections to date have used non-parametric approaches to estimate the noise spectrum around the time of the event~\citep{LIGOScientific:2025zdk,LIGOScientific:2025slb}. The {\textsc{BayesLine}}~\citep{Littenberg:2014oda} algorithm is typically used, which independently fits the noise from each detector by using a two-component non-parametric model composed of a cubic spline to fit the broadband noise spectrum and a linear sum of Lorentzians to fit the narrowband features, such as calibration lines. It uses a reversible jump Markov-Chain Monte-Carlo (RJMCMC) algorithm to determine the number of control points in the spline and the number of Lorentzians to best fit the data. Non-parametric approaches have not been regularly proposed for LISA global fit applications. An exception is {\textsc{GLASS}}~\citep{Littenberg:2023xpl}. {\textsc{GLASS}} uses a simplified version of the {\textsc{BayesLine}} algorithm, which estimates the noise with an Akima spline~\citep{akima1970new}. It does not use a cubic spline as it was found to be prone to oscillations between control points, which was amplified from the large dynamic range and steeply changing spectrum at low frequencies for LISA. Unlike for ground-based GW applications, {\textsc{GLASS}} does not use additional model components to fit features beyond the base spline as a reference noise spectrium that encodes e.g. the TDI transfer function nulls is used.

\section{The GW likelihood} \label{sec:likelihood}

Accurately describing the noise is crucial to find GW signals and avoid biased parameter estimates with misrepresented uncertainties. Similar to inferring the properties of e.g. a massive binary black hole in LISA~\citep[see e.g.][]{Katz:2021uax,Hoy:2023ndx}, a noise model can be fit to the observed data through a process known as Bayesian inference. Although non-Bayesian methods exist, for example Welch's method or maximum likelihood estimates~\citep{Burke:2026aaa}, Bayesian methods are preferred as they provide the full posterior distribution of the parameters rather than a single point estimate, allowing uncertainties and correlations to be quantified consistently.

\begin{figure*}
    \includegraphics[width=0.9\textwidth]{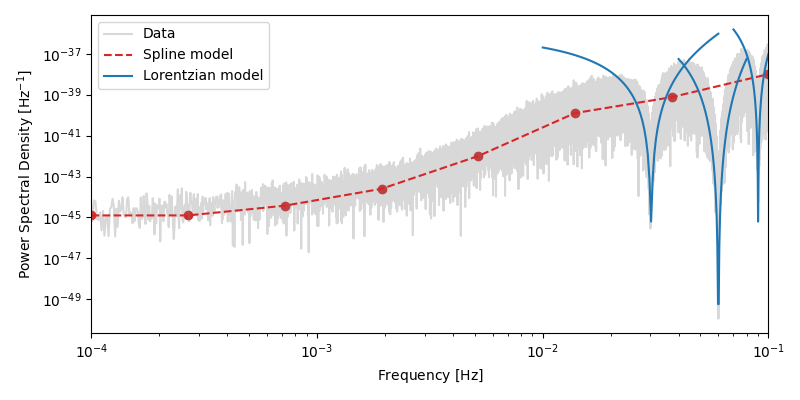}
    \caption{Figure showing our deconstructed model used to estimate the Power Spectral Density of LISA's noise. The red dashed line shows the underlying spline model, the blue lines show the inverse Lorentzian's used to model the loss in sensitivity at high frequencies from the TDI transfer functions, and the grey shows the periodogram for the $X$ TDI channel for reference. In practice the inverse Lorentzians are connected to the Spline with a smooth Sine-squared taper.
    }
    \label{fig:example_spline}
\end{figure*}

In general, Bayesian inference is the process of estimating the parameters $\theta$ from some data $d$ given a model $m$. The parameters are represented by the posterior probability distribution, $P(\boldsymbol{\theta} | d, m)$, and calculated through Bayes theorem,

\begin{equation}
    P(\boldsymbol{\theta} | d, m) = \frac{\mathcal{L}(d | \boldsymbol{\theta}, m)\Pi(\boldsymbol{\theta} | m)}{\zeta},
\end{equation}
where $\mathcal{L}(d | \boldsymbol{\theta}, m)$ is the likelihood, $\Pi(\boldsymbol{\theta} | m)$ is the prior and $\zeta = \int{\mathcal{L}(d | \boldsymbol{\theta}, m)\Pi(\boldsymbol{\theta} | m)\, d\boldsymbol{\theta}}$ is the evidence. Assuming the noise is Gaussian and stationary, the GW log likelihood takes the Whittle form~\citep{wainstein1970extraction,Veitch:2008ur,Veitch:2014wba},

\begin{equation} \label{Eq:likelihood}
\begin{split}
    \log\mathcal{L}(d \mid \boldsymbol{\theta}, m) = & -\frac{1}{2}\langle d - m(\boldsymbol{\theta}) \mid d - m(\boldsymbol{\theta})\rangle \\
    & - \int{\frac{1}{2}\log\big(2\pi S\big)\,df},
\end{split}
\end{equation}
where $\langle a | b \rangle$ is the inner product between two frequency series~\citep{Finn:1992wt},

\begin{equation} \label{Eq:inner_product}
    \langle a \mid b \rangle = 4\mathrm{Re}\int_{0}^{\infty}{\frac{\tilde{a}^{*}(f)\, \tilde{b}(f)}{S(f)}\, df},
\end{equation}
and $S(f)$ is the known one-sided PSD of the noise and $\tilde{a}^{*}$ is the complex conjugate of $\tilde{a}$. In practice, the data are sampled at $N$ discrete frequencies, and the inner product is simplified to,

\begin{equation} \label{Eq:discrete_inner_product}
    \langle a \mid b \rangle = 4\Delta f\,\mathrm{Re}\sum_{k=1}^{N}{\frac{\tilde{a}_{k}^{*}\, \tilde{b}_{k}}{S_{k}}}.
\end{equation}

When the PSD is unknown, it can also be estimated through Bayesian inference~\citep[see e.g.][]{Littenberg:2014oda,Cornish:2014kda}. For example, assuming a noise model $\mathfrak{M}$ with parameters $\boldsymbol{\lambda}$, the posterior distribution for the PSD can be inferred alongside the parameters of the observed GW signal through Bayes theorem,

\begin{equation}
    P(\boldsymbol{\theta}, \boldsymbol{\lambda} | d, m, \mathfrak{M}), = \frac{\mathcal{L}(d | \boldsymbol{\theta}, \boldsymbol{\lambda},  m, \mathfrak{M})\Pi(\boldsymbol{\theta} | m)\Pi(\boldsymbol{\lambda} | \mathfrak{M})}{\zeta'}.
\end{equation}
Although not necessary, we have assumed that the joint prior is independent and can be factorised into individual signal and noise components. The discrete likelihood in Eq.~\ref{Eq:likelihood} becomes,

\begin{equation} \label{Eq:likelihood_noise_model}
\begin{split}
    \log\mathcal{L}(d \mid \boldsymbol{\theta}, \boldsymbol{\lambda}, m, \mathfrak{M}) = & -\frac{1}{2}\langle d - m(\boldsymbol{\theta}) \mid d - m(\boldsymbol{\theta})\rangle \\ 
    & - \Delta f\sum\frac{1}{2}\log\big(2\pi\,\mathfrak{M}(\boldsymbol{\lambda})\big),
\end{split}
\end{equation}
and the discrete inner product in Eq.~\ref{Eq:matrix_inner_product} is,

\begin{equation}
    \langle a \mid b \rangle = 4\Delta f\,\mathrm{Re}\sum_{k=1}^{N}{\frac{\tilde{a}_{k}^{*}\, \tilde{b}_{k}}{\mathfrak{M}_{k}(\boldsymbol{\lambda})}},
\end{equation}

\subsection{The noise covariance matrix}

Although $S$ is often treated as a function of frequency, it is more precisely (up to normalisation) the noise covariance matrix in the frequency domain~\citep[see e.g.][]{Edy:2021par, Burke:2025bun, Burke:2026rjq}. For stationary noise, the different frequency bins are uncorrelated, and the matrix is diagonal and real valued,

\begin{equation}
    \mathbf{S} =
    \begin{pmatrix}
        S_1          & 0            & \cdots & 0            \\
        0            & S_2          & \cdots & 0            \\
        \vdots       & \vdots       & \ddots & \vdots       \\
        0            & 0            & \cdots & S_N
    \end{pmatrix},
    \label{eq:stationary_noise_model}
\end{equation}
where $N$ is the number of frequency bins and $S_{N}(\boldsymbol{\lambda})$ is the PSD at $N$th frequency bin. In this form, Eq.~\ref{Eq:discrete_inner_product} is a matrix multiplication and can be written as,

\begin{equation} \label{Eq:matrix_inner_product}
    \langle a \mid b \rangle = 4\Delta f\,\mathrm{Re}\left(\tilde{\mathbf{a}}^{\dagger}\mathbf{S}^{-1} \tilde{\mathbf{b}}\right),
\end{equation}
where $\tilde{\mathbf{a}} = (\tilde{a}_1, \dots, \tilde{a}_N)^{\mathsf{T}}$ and similarly for $\tilde{\mathbf{b}}$.

When the noise is non-stationary, such that different frequency bins are correlated with one another, the covariance matrix has off-diagonal terms and is complex valued~\citep[with the complex terms encoding information about time, see e.g.][]{Cornish:2026tjt, Burke:2025bun},

\begin{equation}
    \mathbf{S} =
    \begin{pmatrix}
        S_{11} & S_{12} & \cdots & S_{1N} \\
        S_{21} & S_{22} & \cdots & S_{2N} \\
        \vdots & \vdots & \ddots & \vdots \\
        S_{N1} & S_{N2} & \cdots & S_{NN}
    \end{pmatrix},
    \label{eq:nonstationary_noise_model}
\end{equation}
where $S_{kk'}$ is the covariance between the $k$th and $k'$th frequency bins.

Up until now, we have assumed a single GW detector. For a network of $N_{\rm d}$ detectors with correlated noise, each element of Eq.~\ref{eq:nonstationary_noise_model} becomes an $N_{\rm d}\times N_{\rm d}$ matrix. In general,
\begin{equation}
    S_{kk'} =
    \begin{pmatrix}
        S_{kk'}^{11}         & S_{kk'}^{12}         & \cdots & S_{kk'}^{1N_{\rm d}}         \\
        S_{kk'}^{21}         & S_{kk'}^{22}         & \cdots & S_{kk'}^{2N_{\rm d}}         \\
        \vdots               & \vdots               & \ddots & \vdots                       \\
        S_{kk'}^{N_{\rm d}1} & S_{kk'}^{N_{\rm d}2} & \cdots & S_{kk'}^{N_{\rm d}N_{\rm d}}
    \end{pmatrix},
    \label{eq:multidetector_noise_model}
\end{equation}
where $S_{kk'}^{IJ}$ is the covariance between the $k$th frequency bin of detector $I$ and the $k'$th frequency bin of detector $J$, with $S_{k'k}^{JI} = \big(S_{kk'}^{IJ}\big)^{*}$. For LISA, $N_{\rm d} = 3$, representing the 3 TDI observables. From now on, we will refer to the network of detectors $N_{\rm d}$, as the three TDI channels since we are focusing on the LISA observatory in this work.

If the noise is assumed to be jointly stationary across the three TDI channels, different frequency bins are uncorrelated and only the blocks with $k=k'$ are non-zero, with correlations between channels at the same frequency retained. $S_{kk}^{II}$ is then the PSD of the individual detector $I$, and the off-diagonal elements $S_{kk}^{IJ}$ are the CSDs between detectors $I$ and $J$. In this work, we generate non-stationary data, with correlated noise properties across the different TDI channels. However, we assume that the noise is locally non-stationary in our analyses, simplifying the noise covariance matrix. The validity of this assumption was explored in~\cite{Burke:2026aaa}. We leave the full non-stationary extension for future work.

\subsection{Stochastic sampling}

Although the GW likelihood is well known, it is difficult to analytically compute the posterior distribution. As such, stochastic sampling algorithms, such as nested sampling~\citep{Skilling:2004pqw, Skilling:2006gxv} and Markov-Chain Monte-Carlo~\citep{metropolis1949monte}, are often used to draw samples from the unknown posterior distribution; although see~\citep{Pankow:2015cra,Lange:2018pyp,Gabbard:2019rde,Green:2020dnx,Green:2020hst,Dax:2021tsq,Tiwari:2023mzf,Fairhurst:2023idl,Dax:2024mcn,Raymond:2024xzj,Roulet:2024hwz,Mushkin:2025yks,Williams:2025szm,Roulet:2026mzz} for alternative approaches.

Unfortunately, classical stochastic samplers scale poorly with dimension. This can make it difficult to sufficiently explore high-dimensional distributions, especially with narrow, multimodal or strongly correlated posterior distributions. As a result, sampling over the full correlated likelihood in Eq.~\ref{Eq:likelihood_noise_model} is difficult without rigorous sampler settings; for example, assuming each term in the noise covariance matrix is modelled with 20 parameters, the full likelihood involves sampling over a 120 dimensional parameter space. One way to improve sampler convergence is to reduce the dimensionality of the likelihood. This can be achieved by considering a blocked approach.

To reduce the computational challenge of stochastically sampling a large parameter space, we can approximate Eq.~\ref{Eq:likelihood_noise_model} by employing a pairwise likelihood, in which the noise covariance matrix is split into several $2\times2$ blocks. For three TDI channels, $X$, $Y$ and $Z$, the $3\times 3$ noise covariance matrix can be replaced by 3, $2\times 2$ matrices,

\begin{equation}
    \begin{pmatrix}
        S^{XX} & S^{XY} \\
        S^{YX} & S^{YY} \\
    \end{pmatrix},
    \begin{pmatrix}
        S^{XX} & S^{XZ} \\
        S^{ZX} & S^{ZZ} \\
    \end{pmatrix},
    \begin{pmatrix}
        S^{YY} & S^{YZ} \\
        S^{ZY} & S^{ZZ} \\
    \end{pmatrix},
\end{equation}
which can be solved in parallel. The computational cost can be further reduced by first inferring the PSDs of the individual channels, $S^{XX}$, $S^{YY}$ and $S^{ZZ}$, independently and in parallel. The CSD for each pair $(I, J)$ can then inferred with the PSDs $S^{II}$ and $S^{JJ}$ fixed at their best fit values. For three TDI channels, $3$ analyses can be launched in parallel to model the PSDs, and once complete, a further $3$ analyses can be launched in parallel to model the CSDs.

This blocked approach significantly reduces the size and degeneracies in the parameter space, as each analysis is now sampling over a e.g. 20 dimensional parameter space. This significantly improves sampler convergence and reduces the wall-time of analyses. This is at the cost of only providing a single estimate for the PSDs of each detector, and therefore the uncertainties on these measurements are neglected when estimating the CSDs. Also, by using this blocked approach, information about e.g. $S^{XY}$ is neglected when considering $S^{XZ}$, and positive definiteness of the full $3\times3$ matrix is not enforced.

To check when the blocked approximation breaks down, we explicitly compared the blocked approach against the full likelihood for different combinations. We found that, as expected, when at most one pair of channels is coherent, the blocked likelihood matches the full likelihood exactly. When two or more pairs are coherent, as is the case for the $X$, $Y$ and $Z$ channels, the blocked likelihood diverges from the full likelihood. By fitting each cross-spectrum independently, the blocked likelihood neglects the coupling between the coherencies, and the discrepancy grows as the coherency increases. Nevertheless, we believe the blocked approach has merit for modelling the noise in LISA. For example, the blocked approximation holds true and will significantly reduce the computational cost when analysing the 
$A$, $E$ and $T$ channels~\citep[which is common, see e.g.][]{Katz:2021uax,Buscicchio:2021dph,Pratten:2022kug,Weaving:2023fji,Hoy:2023ndx,CabournDavies:2024hea,Katz:2024oqg,Sharma:2024sfb}, which are quasi-orthogonal in second generation TDI and therefore have small coherences. 

\subsection{The PSD model} \label{sec:psd_model}

Similar to ground-based GW applications, we propose using a two-component model to fit the PSDs of the LISA observatory: one component to fit the broadband noise spectrum and another to fit high frequency regions where the noise formally goes to zero due to the TDI transfer functions. The latter is crucial due to the geometry of the LISA detector: the noise drops to exactly zero for signals with wavelength that fit an integer number of cycles within the detector arm.

We model the broadband noise spectrum with an Akima spline, $g(f)$, and the high frequency features with a linear combination of inverse Lorenztians. The Akima spline is controlled by $N + N_{g}$ knots, where $N$ is the number of knots within the frequency band of interest and $N_{g}$ are a limited number of \emph{ghost} knots to control edge effects. 

In second generation TDI, the nulls in the TDI transfer function occur due to a $\sin^{2}(2\pi fL/c)\sin^{2}(4\pi fL/c)$ term in the transfer function~\citep{armstrong1999time,Estabrook:2000ef,Vallisneri:2005ji,Tinto:2020fcc,tinto2023second}, where $L$ is the arm length of the observatory. At odd multiples of $30\,\mathrm{mHz}$ only the $\sin^{2}(4\pi fL/c)$ term vanishes, whereas at multiples of $60\,\mathrm{mHz}$ both terms vanish, producing a deeper and sharper feature; a fourth order zero for the $60\,\mathrm{mHz}$ null compared to a second order zero at the $30\,\mathrm{mHz}$ and $90\,\mathrm{mHz}$ nulls. Owing to this, we use two different inverse Lorenztian models for the different structures: a locally quadratic model for the $30\,\mathrm{mHz}$ and $90\,\mathrm{mHz}$ nulls, and a locally quartic model for the $60\,\mathrm{mHz}$ null. In all cases, the inverse Lorentzian, $L(f)$, is parametrised with a central frequency $f_{0}$, a quality factor $\gamma$, and a minimum value at the resonant frequency $S_{\mathrm{min}}$, 

\begin{equation}
    L(f) = g(f) - \frac{g(f) - \textrm{S}_{\mathrm{min}}}{1 + \left(\frac{f - f_{0}}{\gamma}\right)^{\mathfrak{N}}},
\end{equation}
where $\mathfrak{N} = 2$ for the $30\,\mathrm{mHz}$ and $90\,\mathrm{mHz}$ nulls, and $\mathfrak{N} = 4$ for the $60\,\mathrm{mHz}$ null.

We show an example of our two-component model in Fig.~\ref{fig:example_spline}. The full PSD is the sum of the base Akima spline and $\mathcal{N}$ inverse Lorentzian fits where $\mathcal{N}=3$ in this work due to our chosen high frequency cut-off. To smoothly transition between the base Akima spline and the inverse Lorentzians, we apply a smooth Sine-squared taper, enabling continuity at the connection frequencies $f_{\mathrm{low}}$ and $f_{\mathrm{high}}$.

\subsection{The CSD model}

For the CSDs, we assume the cross spectrum is a combination of real and imaginary components. Rather than fitting $S^{IJ}_{kk'}$ directly, as in the PSD case, we model $S^{IJ}_{kk'} = \rho_{IJ}\,\sqrt{S_{II} S_{JJ}}$ where $\rho_{IJ}$ is defined as the complex coherency between detectors $I$ and $J$. Under this parameterisation, $\rho_{IJ}$ quantifies how strongly the two detectors are linked, and ranges between $0$, uncorrelated, to 1, perfectly correlated. For the model to be valid, $|\rho_{IJ}| < 1$.

\begin{figure*}
    \includegraphics[width=0.98\textwidth]{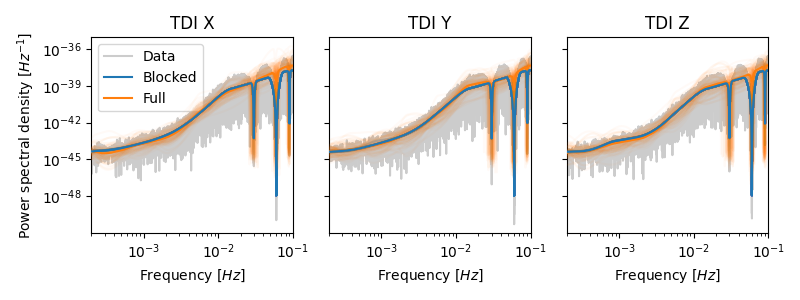}
    \caption{Plot comparing the periodogram for the TDI $X$, $Y$ and $Z$ (grey) with the PSDs inferred using a sum of an Akima spline and 3 inverse Lorentzians (we can divide through by the laser central frequency to obtain the familiar $\mathrm{Hz}^{-1}$ units). We show the PSDs inferred according to the Blocked likelihood (blue), where the noise covariance matrix is split and solved in parallel, the full likelihood (orange), where the PSDs and CSDs are solved in a single Bayesian analysis, and a reference Welch estimate (green) based on Hann-windowed 24 hour segments with 50\% overlap. The solid thick lines show the median estimate, and the thin lines show 20 random draws from the inferred posterior.
    }
    \label{fig:psds}
\end{figure*}

To account for both the real and imaginary components while enforcing $|\rho_{IJ}| < 1$, we define $\rho_{IJ} = \kappa r$, where
$r = g_{\mathrm{re}} + i g_{\mathrm{im}}$ and,

\begin{equation}
    \kappa = \begin{cases}
        \dfrac{\tanh|r|}{|r|}, & |r| > 0,\\[4pt]
        1, & |r| = 0,
    \end{cases}
\end{equation}
In this form, we are free to model the real and imaginary components of $r$ independently with individual Akima splines.

In~\cite{Burke:2026aaa}, we demonstrated that unbiased and accurate noise inference can be obtained for the CSDs when modelling only the real component; neglecting the imaginary components widen the knot uncertainties by at most $3\%$ even at the most asymmetric LISA constellation in a two year period. We nevertheless model both the real and imaginary components in this work to stress test the CSD model.

\section{Results} \label{sec:results}

\subsection{Analysis setup}

To verify the non-parametric models described in Section~\ref{sec:models}, and to test the different likelihood approaches described in Section~\ref{sec:likelihood}, we analysed 30 days of data with realistic LISA orbits, split into 7 day segments. We inferred the noise model through Bayesian inference between the frequencies $10^{-4}\,\mathrm{Hz} < f < 10^{-1}\, \mathrm{Hz}$. The simulation parameters for this dataset are nearly identical to those listed in Table I of~\cite{Burke:2026aaa}. We generated a noise-only dataset generated with \texttt{lisainstrument}~\citep{bayle2023lisainstrument,Bayle:2022okx} using ESA trailing orbits described in \texttt{lisaorbits}~\citep{Bayle_lisaorbits}, and converted to the TDI observables using extended precision via \texttt{pytdi}~\citep{Staab2022PyTDI}. The data was recast using the \texttt{mojito} package~\citep{mojito_2026} and post-processed using \texttt{MojitoProcessor}~\citep{Burke_MojitoProcessor_2026}. The only difference with~\cite{Burke:2026aaa}, is that we elected not to downsample the data to a cadence of $\Delta t = 5\,\mathrm{s}$, as we wanted to explicitly show the efficacy of our PSD and CSD models for the sharper null at $60\,\mathrm{mHz}$, which would otherwise be removed as a result of downsampling~\citep{Burke:2026aaa}. We analyse 7 days of data so we can safely assume stationarity, and hence validate our assumption that the likelihood is described by a Whittle function. In 1 day of data, LISA's arm length is expected to vary by at most 0.035\%~\citep{Hartig:2024ojo}, meaning $< 0.25\%$ in 7 days of data.

To sample the posterior distribution, we used Markov-Chain Monte-Carlo (MCMC) methods. We specifically used the \texttt{Eryn}~\citep{Karnesis:2023ras} sampler since it allows for Reversible-Jump parallel-tempered MCMC, where the dimensionality of the model can change during the sampling (Reversible-Jump), and multiple tempered chains can be launched in parallel to avoid local maxima (parallel-tempering). This is useful for e.g. dynamically choosing the number of knots in the base Akima splines of our models. We performed a small convergence study on the 1 day of stationary Gaussian noise coloured by the \texttt{SciRDv1} noise PSD~\citep{Babak:2021mhe} to identify sufficient sampler settings for robust results in a large prior volume. We settled on $800$ walkers, $8$ parallel temperatures, $2000$ steps, and $5000$ samples discarded as burn-in. These settings are comparable to those used in previous works with the \texttt{Eryn} sampler~\citep{Puecher:2026poi}. We found that these settings reduced the variance in median inferred PSD, as well as reduced the inferred 90\% credible interval when using the blocked likelihood. We leave a detailed sampler convergence study for the full and blocked likelihoods for future work. Unless otherwise stated, we seeded the MCMC walkers with a random point chosen from the prior. 

\begin{figure}
    \includegraphics[width=0.47\textwidth]{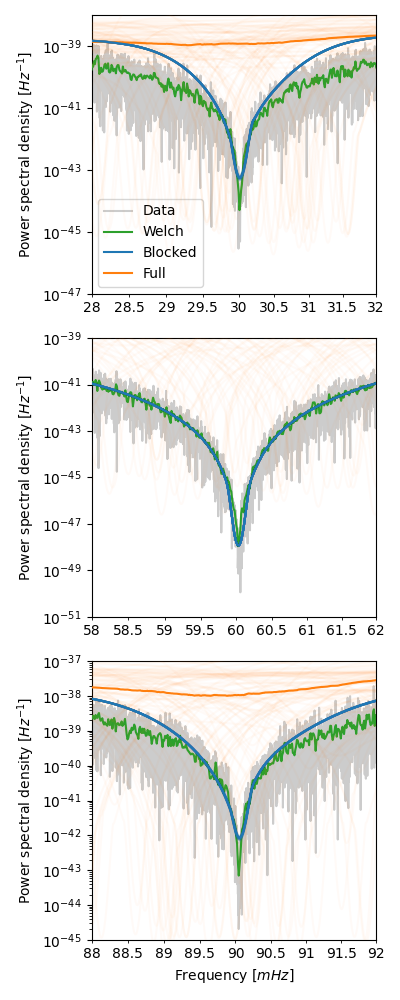}
    \caption{Plot comparing the periodogram for the TDI $X$, $Y$ and $Z$ (grey) with the PSDs inferred using a sum of an Akima spline and 3 inverse Lorentzians at each of the transfer function nulls. We only show the PSDs inferred according to the Blocked likelihood (blue), where the noise covariance matrix is split and solved in parallel. The solid thick lines show the median estimate, and the thin lines show 20 random draws from the inferred posterior. 
    }
    \label{fig:psds_zoom}
\end{figure}

For both the PSD and CSD models described in Section~\ref{sec:models}, we placed 8 knots uniformly in log frequency within the window: $0.05\,\mathrm{mHz} < f < 100\,\mathrm{mHz}$, and 6 ghosts knots, 3 either side of the window placed between $0.001\,\mathrm{mHz} < f < 0.05\,\mathrm{mHz}$ and $100\,\mathrm{mHz} < f < 1\,\mathrm{Hz}$ respectively to control edge effects. The number of knots were chosen based on a preliminary investigation, where an RJMCMC analysis was performed on the broadband noise spectrum consisting of only the analytic TMA and OMS noise. When allowing the number of knots to vary, we found a small preference for 8 knots. Future work will allow the number of knots to vary during each MCMC analysis. In both the PSD and CSD model, we assumed a log-uniform distribution for the knot amplitudes between $10^{-47}\,\mathrm{Hz}^{-1} < \mathcal{A} < 10^{-37}\,\mathrm{Hz}^{-1}$. Finally, for the 3 inverse Lorentzians used in the PSD model, we assumed a uniform prior for the central frequency $f_{0}$, and uniform distributions for the lower and upper connection frequencies $f_{\mathrm{low}}$ and $f_{\mathrm{high}}$ to the base spline. Owing to the different functional forms of the inverse Lorentzians, we employed different priors for the quality factors: we assumed a uniform distribution between $10$ and $30$ for the $30$ and $90\,\mathrm{mHz}$ nulls and a uniform distribution between $2$ and $30$ for the $60\,\mathrm{mHz}$ null. In our analyses, we do not allow for the ``lack of TDI null'' case, where the quality factor is $0$ by definition. We purposely use wide priors for the knot amplitudes to remain as agnostic as possible. Smaller and more tailored priors could be used for each knot amplitude, for example based on a Welch estimate, however, this would mean that the same data is used twice: once for informing the prior, and once for obtaining the posterior. Posterior repartitioning~\citep{chen2019improving,Prathaban:2026kft} offers a solution to using the same data twice, but this is not explored in this work.

\subsection{Inferred PSDs and CSDs}

In Figure~\ref{fig:psds} we compare the inferred PSDs with the periodogram for the TDI $X$, $Y$ and $Z$ observables. When focusing on the median inferred PSD, we see that for all three channels, the blocked likelihood more closely matches the underlying periodogram: while the broadband structure of the individual PSDs from both methods are similar, the median estimate from the blocked likelihood more faithfully captures the nulls in the PSD due to the TDI transfer functions. This can be seen more clearly in Figure~\ref{fig:psds_zoom}.

\begin{figure*}
    \includegraphics[width=0.98\textwidth]{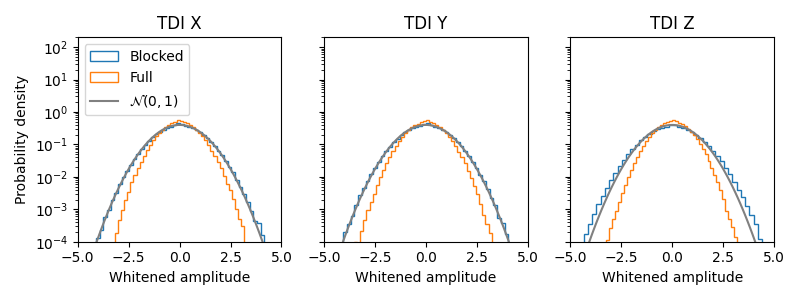}
    \caption{Plot comparing whitened LISA data to the expected Normal distribution for the TDI $X$, $Y$ and $Z$ channels. In blue (orange) we show distribution of whitenened amplitudes for the PSD inferred with the blocked (full) likelihood approach. In both cases, the data was whitened by the median inferred PSD assuming the non-parametric noise model described in Section~\ref{sec:psd_model}.
    }
    \label{fig:scirdv1_whitened_data}
\end{figure*}

\begin{figure*}
    \includegraphics[width=0.98\textwidth]{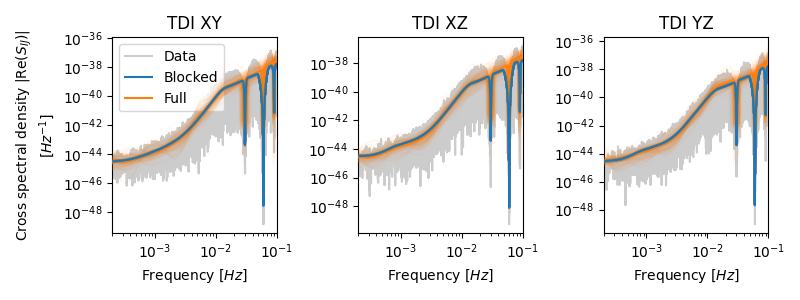}
    \caption{Plot showing the CSDs inferred using an Akima spline for the complex coherence for the TDI $XY$, $XZ$ and $YZ$ observables. We compare the cross periodograms (grey) against the CSDs inferred according to the Blocked likelihood (blue), where the noise covariance matrix is split and solved in parallel, and the full likelihood (orange), where the PSDs and CSDs are solved in a single Bayesian analysis. The solid thick lines show the median estimate, and the thin lines show 20 random draws from the inferred posterior.
    }
    \label{fig:csds}
\end{figure*}

We believe the differences between the median inferred PSDs from the two likelihood approaches are due to poor sampler convergence when using the full likelihood. We found that the sampler struggled to find viable solutions, which is reflected in the large spread of individual PSD draws in Figures~\ref{fig:psds} and \ref{fig:psds_zoom}. We see that for some draws, the nulls are captured well, but due to large uncertainties, the median estimate averages over the nulls to a smooth function. The sampler diagnostics support this interpretation. For the blocked likelihood, consecutive samples show the autocorrelation typical of an MCMC chain, where each step is a small move from the previous position. For the full likelihood, consecutive samples are effectively uncorrelated, and every proposed swap between temperatures is accepted. This indicates that the full likelihood provides little information to guide the walkers towards the high-probability region of the posterior.

The full likelihood is sampling over a 162 dimensional parameter space -- $26$ for each PSD (14 for the spline amplitudes and 12 for the three inverse Lorentzians) and 28 for each CSD (14 for both the real and imaginary components -- while the blocked likelihood is sampling over at most $28$ dimensions in a single analysis. The significant increase in the dimensionality, while keeping the sampler settings consistent, means the sampler explores a much smaller fraction of the posterior volume for the same number of samples when using the full likelihood. Scaling the autocorrelation time measured for the full likelihood with the number of dimensions, we estimate that \texttt{Eryn} would require a burn-in of $> 2 \times 10^4$ steps, followed by $> 5 \times 10^4$ steps of sampling, to obtain robust results; an order of magnitude more than used here. This highlights that alternative sampling approaches~\citep[see e.g.][]{williams2026inprep} may be required if the full likelihood is used due to the significant increase in computational cost.

Looking specifically at Figure~\ref{fig:psds_zoom}, we see that the combination of our PSD model and the blocked likelihood generally fits the nulls in the transfer function well. Despite wide priors, we find that the median inferred central frequency of each null is found within at most $0.3\%$, with the $60\,\mathrm{mHz}$ model performing optimally: $0.3\%$, $0.05\%$ and $0.3\%$ for the $30\,\mathrm{mHz}$, $60\,\mathrm{mHz}$ and $90\,\mathrm{mHz}$ nulls respectively. Interestingly, we find that the tails of the inverse Lorentzian for the $30\,\mathrm{mHz}$ result slightly overestimate the noise when using the blocked likelihood. We found that this was primarily influenced by the inferred $60\,\mathrm{mHz}$ null: if narrow priors are chosen such that the $60\,\mathrm{mHz}$ null is misrepresented, the $30\,\mathrm{mHz}$ and $90\,\mathrm{mHz}$ perfectly describe the data. This implies that it may not be a problem when downsampling the data to a cadence of $\Delta t = 5\,\mathrm{s}$, where the $60\,\mathrm{mHz}$ null is removed due to downsampling~\citep{Burke:2026aaa}. Despite large uncertainties, with individual PSD draws varying significantly across each null, we found that the full likelihood result still infers median central frequencies of each null to within at most $1\%$ of the truth. In general, we also see good agreement with a reference Welch estimate, both across the broadband and for each null, providing further confidence that our non-parametric fit is performing as expected.

In Figure~\ref{fig:scirdv1_whitened_data}, we assess how well each method whitens the data in each TDI channel. In both cases, we whitened by the median inferred PSD. If our inferred PSD is a good estimate of LISA's noise, we expect the distribution of the whitened amplitudes should resemble a unit normal distribution (assuming the data is Gaussian and stationary). Agreement with a unit normal is a necessary, though not sufficient, indication of a good PSD estimate. As can be seen, we obtain the expected normal distribution for the blocked analysis. We see a small divergence from the expected normal distribution in the tails for the Z channel, but this will likely be resolved with stricter sampler settings. Since the median PSD inferred by the full likelihood is failing to capture the nulls due to poor convergence, we obtain a distribution with a smaller than expected unit variance. This highlights that it is not not whitening the data as well as the blocked approach.

Finally, in Figure~\ref{fig:csds} we compare the inferred CSDs with the cross periodograms for the TDI $XY$, $XZ$ and $YZ$ observables. We see a similar story as in Figure~\ref{fig:psds}: the blocked likelihood more closely matches the cross periodograms. We see that the full likelihood obtains larger uncertainties, which is likely because the blocked likelihood is not accounting for errors on the PSDs. Nevertheless, if only a median estimate is used, consistent with ground-based GW detectors~\citep{LIGOScientific:2025zdk}, the blocked likelihood more closely resembles the periodogram.

\section{Discussion} \label{sec:discussion}

Modelling the noise distribution in the LISA observatory is crucial in order to search for GW signals, infer unbiased source properties, and maximise the scientific return of the LISA mission. Although simple methods exist~\citep{Welch:1967oth} that are commonly used throughout astronomy, more complex models are needed for LISA since the data will be non-stationary, with overlapping signals, and no noise-only segments of off-source data are available. In this work, we introduced a non-parametric model to fit the full and complex noise covariance matrix, with a combination of splines and inverse Lorentzians. Our model is completely agnostic and requires no prior knowledge of the LISA detector or the TDI transfer functions. 

By analysing 7 days of simulated noise-only LISA data with realistic orbits, we demonstrated that our model fits the periodograms of the different TDI observables to good precision and whitens the data as expected. When comparing to Welch estimates we see good agreement, which is expected since our dataset contains no overlapping signals and any effects from non-stationarity is minimal. We emphasise that for realistic LISA scenarios, our model offers advantages beyond Welch since it can estimate the noise distribution on-source, and can be included within a global fit framework to jointly infer the astrophysical source properties and the noise model. However, we note that a fully model agnostic approach may be degenerate with LISA's stochastic background~\citep{Adams:2010vc}.

We also considered two approaches to obtain the full noise convariance matrix: one where the full likelihood is split and solved in parallel with reduced dimensionality, and one where the full likelihood is solved in a single analysis. We found that although the split likelihood is an approximation that breaks down in highly correlated data, it is the only approach that gave converged results. We expect the full likelihood to give consistent results when significantly stricter settings are used. However, we note that in the majority of previous work~\citep{Katz:2021uax,Buscicchio:2021dph,Pratten:2022kug,Weaving:2023fji,Hoy:2023ndx,CabournDavies:2024hea,Katz:2024oqg,Sharma:2024sfb}, only the PSDs are used when modelling the noise, and the blocked approach provides robust and converged results for this case.

We envision that our non-parametric noise model can be used as part of the global fit for estimating the noise on-source, or used in conjunction with modelled approaches to identify, quantify, and constrain the frequency dependence for unknown noise artifacts once LISA flies.

\section{Acknowledgments}

We thank Sophie Bini, Neil Cornish, Peter Lott, Andrew Lundgren, Joe Metson, Julianna Ostrovska and Saulo Soares-de-Albuquerque-Filho for PSD related discussions in the context of ground-based GW observatories, which were helpful for this work. We also thank the UK LISA ground segment for comments on this work. C.H thanks the University of Portsmouth for support through the Dennis Sciama Fellowship. J.B. and O.B. acknowledge financial support from the Grant UKRI972 awarded via the UK Space Agency, and L.K.N thanks the UKRI Future Leaders Fellowship for support through the grant UKRI2745. We are also grateful for computational resources provided by the SCIAMA high performance computing cluster which is supported by the Institute of Cosmology and Gravitation (ICG) and the University of Portsmouth. For the purpose of open access, the author(s) has applied a Creative Commons Attribution (CC BY) licence to any Author Accepted Manuscript version arising.

\section{Data Availability}

The analysis code behind this paper, including the non-parametric models for the PSDs and CSDs, is available through the \texttt{lisa-noise} package at \href{https://doi.org/\lisanoisedoi}{DOI \lisanoisedoi}.

\bibliographystyle{mnras}
\bibliography{refs}
\bsp 
\label{lastpage}
\end{document}